\documentclass[prl,showpacs,twocolumn,superscriptaddress,aps]{revtex4-1}
\usepackage[utf8]{inputenc}
\usepackage[american,british]{babel}
\usepackage[T1]{fontenc}
\usepackage[pdftex]{graphicx}  
\usepackage{graphicx, xcolor}
\usepackage{dcolumn}
\usepackage{bm}
\usepackage{amsmath,amsthm,amssymb}
\usepackage{color}
\usepackage{verbatim}
\usepackage{ulem}

\definecolor{darkGreen}{RGB}{0,110,0}
\definecolor{darkBlue}{RGB}{0,0,130}
\usepackage[colorlinks,citecolor=darkGreen,linkcolor=darkBlue,urlcolor=blue,hyperindex]{hyperref}

\begin{document}
\title{Breakdown of ergodicity in disordered $U(1)$ lattice gauge theories}
\author{G. Giudici}
\affiliation{The Abdus Salam International Center for Theoretical Physics, Strada Costiera  11,  34151  Trieste,  Italy}
\affiliation{SISSA, via Bonomea 265, 34136 Trieste, Italy}
\author{F. M. Surace}
\affiliation{The Abdus Salam International Center for Theoretical Physics, Strada Costiera  11,  34151  Trieste,  Italy}
\affiliation{SISSA, via Bonomea 265, 34136 Trieste, Italy}
\author{J. E. Ebot}
\affiliation{The Abdus Salam International Center for Theoretical Physics, Strada Costiera  11,  34151  Trieste,  Italy}
\author{A. Scardicchio}
\affiliation{The Abdus Salam International Center for Theoretical Physics, Strada Costiera  11,  34151  Trieste,  Italy}
\affiliation{INFN Sezione di Trieste, Via Valerio 2, 34127 Trieste, Italy}
\author{M. Dalmonte}
\affiliation{The Abdus Salam International Center for Theoretical Physics, Strada Costiera  11,  34151  Trieste,  Italy}
\affiliation{SISSA, via Bonomea 265, 34136 Trieste, Italy}

\date{\today}


\begin{abstract}
We show how $U(1)$ lattice gauge theories display key signatures of ergodicity breaking in the presence of a random charge background. Contrary to the widely studied case of spin models, in the presence of Coulomb interactions, the spectral properties of such lattice gauge theories are very weakly affected by finite-volume effects. This allows to draw a sharp boundary for the ergodic regime, and thus the breakdown of quantum chaos for sufficiently strong gauge couplings, at the system sizes accessible via exact diagonalization. Our conclusions are independent on the value of a background topological angle, and are contrasted with a gauge theory with truncated Hilbert space, where instead we observe very strong finite-volume effects akin to those observed in spin chains.
\end{abstract} 

\maketitle

\paragraph{Introduction. -- } Ergodicity is one of the pillars of statistical mechanics. In the quantum regime, the ergodic hypothesis and the corresponding eigenstate thermalization hypothesis (ETH) \cite{srednicki1994chaos,deutsch1991quantum} provide a sensible justification for the use of microcanonical and canonical ensembles {\it in lieu} of their Hamiltonian dynamics to compute long term averages of observables. An established mechanism to circumvent thermalization is provided by Anderson localization~\cite{anderson1958absence}. The latter describes how non-interacting systems can feature a dynamical phase in which diffusion (and hence transport) and ergodicity are suppressed without any need to fine-tune the Hamiltonian to an integrable one. Remarkably, this mechanism has been shown to survive the introduction of interactions at the perturbative level~\cite{Basko:2006hh,gornyi2005interacting}, a phenomenon dubbed many-body localization (MBL) ~\cite{altman2015review,huse2015review,imbrie2017review,abanin2019colloquium}. However, owing to the fundamentally more complex nature of many-body theories, establishing the breakdown of ergodicity and characterizing the ergodic/non-ergodic transition in generic, interacting microscopic models has proven challenging. At the practical level, this is due to the fact that quantum chaos (which underlies ETH) is ultimately linked to the full spectral content of a theory \cite{haake1991quantum}, where the applicability of analytical techniques is less established with respect to low-energy studies~\cite{Fleishman1980interactions,altshuler1997quasiparticle,Basko:2006hh,gornyi2005interacting,imbrie2014many,ros2015integrals,de2014asymptotic,Roeck2014:Scenario,de2016absence,Chandran2016beyond}.

An archetypal example in this field has been the one-dimensional (1D) Heisenberg model with random fields~\cite{Pal2010}, where, in the absence of SU(2) symmetry~\cite{prelovvsek2016absence,potter2016symmetry,protopopov2017effect,protopopov2019non}, first signatures of the breakdown of ergodicity were established at finite volume. 
Despite a follow-up impressive numerical effort~\cite{de2013ergodicity,John2015TotalCorrelations,alet2015,pietracaprina2016entanglement,Znidaric2016Diffusive,alet2018many}, the precise location of the localization transition in this and similar microscopic models is still actively debated. A systematic drift (of order 1 in units of the disorder strength) of the would-be critical disorder strength was noticed already as early as in Ref.~\cite{Pal2010}. 
The finite-size scaling theory close to the phase transition is also still far from being satisfactory, with the numerically extracted critical exponents~\cite{alet2015,pietracaprina2016entanglement,mace2019multifractal} at odds with strong disorder renormalization group predictions \cite{AltmanTheory2015,dumitrescu2019kosterlitz}, and not compatible with the Harris criterion \cite{chandran2015finite,Harris_1974}. A recent analysis based on a different finite-size scaling ansatz was proposed where the transition point drifts {\it linearly} with system sizes~\cite{untajs2019quantum}, which however seems to apply, at small sizes, also to models where localization is demonstrated on solid grounds~\cite{sierant2019,abanin2019reply}. 
On top of this, a recent analysis discussed how large a system size one should analyze to go beyond the transient behavior in numerical or experimental studies~\cite{p2019study}. The challenge is thus to identify generic mechanisms, and concrete model Hamiltonians, that allow to determine the breakdown of ergodicity at system sizes accessible to numerical simulations and experiments~\cite{Bloch2015,monroe2015}.

\begin{figure}
{\includegraphics[scale=0.53]{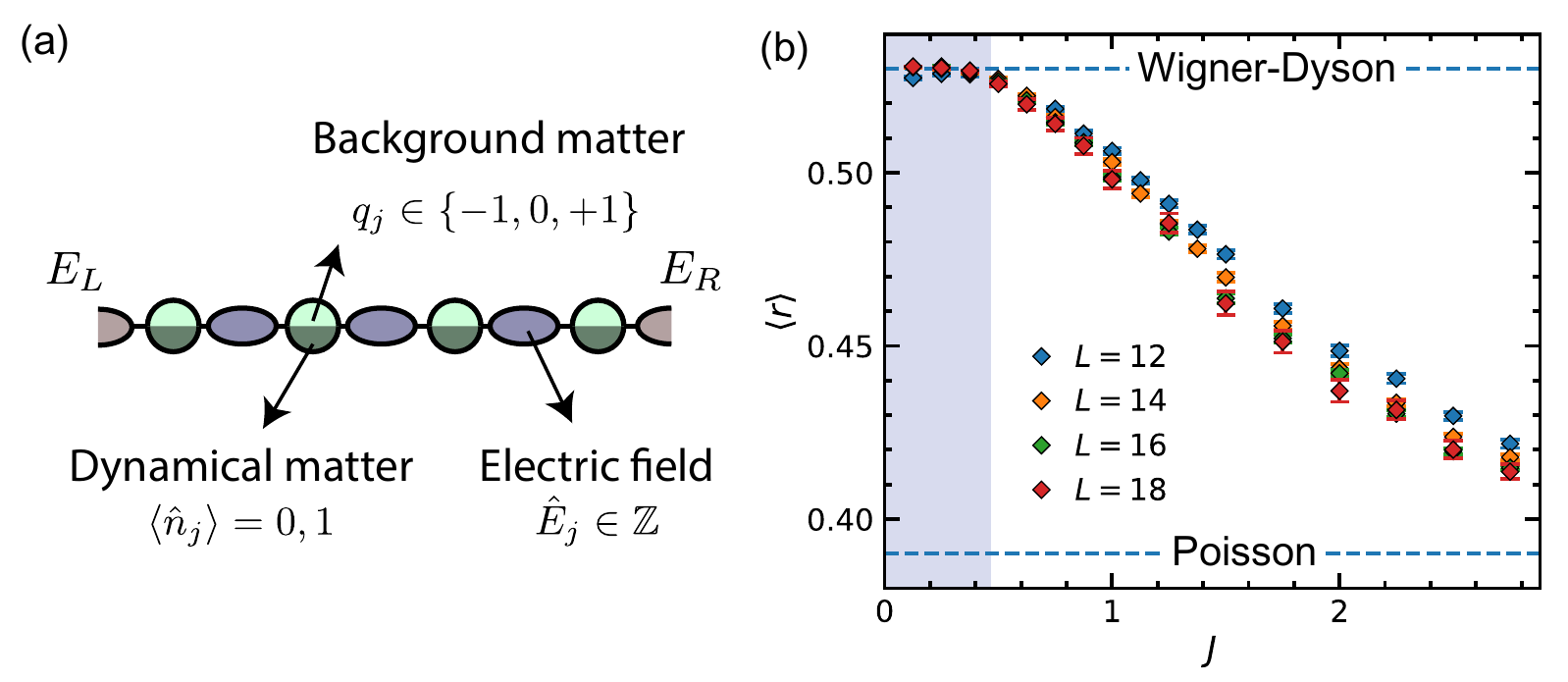}}
\caption{(a) Schematics of U(1) lattice gauge theories. The $U(1)$ gauge field lives on link between the sites of the chain. Dynamical matter (dark green) is a fermionic variable living on the sites, while static charges (light green) are random integers which take values $0,\pm1$. (b) Average level spacing (see Eq.\,\eqref{r_ratio}) as a function of the gauge coupling $J$ for different $L$ (see text). The shaded region represents the estimated ergodic phase. }
\label{fig1} 
\end{figure}

In this work, we show how lattice gauge theories (LGTs)~\cite{wilson1974confinement,Montvay1994} provide a framework within which the transition between ergodic and non-ergodic behavior can be studied using conventional, well controlled numerical methods. The key element of this observation is the {\it cooperative} effect of disorder and Coulomb law, which leads to a localization phenomenon that - as we show below - is parametrically different from what observed in other models. In concrete, we illustrate this mechanism in the context of the 1D lattice Schwinger model - quantum electrodynamics in 1D, illustrated schematically in Fig.~\ref{fig1}(a). A sample of our results is depicted in Fig.~\ref{fig1}(b), which shows one of the most popular witnesses of ergodicity, i.e.\ the averaged ratio of nearby gaps, as a function of the gauge coupling. The results display a sharp departure from Wigner-Dyson expectations, and, crucially, the transition point from the corresponding plateau is unaffected by finite-volume effects. This behavior reflects into a modified functional form of the spectral form factor, which is not compatible with ergodicity.

Before entering into the details of our treatment, we find useful to illustrate a qualitative reasoning why 1D gauge theories may be an ideal candidate to display a somehow smoother behavior in terms of finite volume effects, and a clearer breakdown of ergodicity. As already mentioned, contrary to typical inter-spin interactions, gauge-field mediated interactions typically slow down the system dynamics~\cite{K_hn_2015, Pichler_2016, brenes2018}, and thus do not necessarily compete with disorder. In the typical Basko-Aleiner-Altshuler (BAA) scenario~\cite{Basko:2006hh}, interactions open up channels for delocalization by allowing a series of local rearrangements to create a resonance between two quantum states. This leads to a competition between disorder (increasing energy differences and denominators in perturbation theory) and interactions (increasing matrix elements and therefore the numerator). In the presence of one-dimensional Coulomb law, interactions cannot be introduced perturbatively and therefore a BAA-like analysis does not work. This is because a local rearrangement of the degrees of freedom (spins or particle occupation numbers) leads to a large (even extensive) change in energy, therefore suppressing the amplitude of having a resonant process. This behavior is unrelated to the case of non-confining long range interactions (e.g.\ which decay like $1/r^\alpha$, see Ref.~\cite{burin2006energy,yao2014many,burin2015localization}), and is reflected in finite-volume properties observed in previous numerical studies~\cite{Nandkishore_2017,brenes2018,Sondhi2018}, that focused on quench dynamcis and local observables. 

\paragraph{Model Hamiltonian. --}

We focus here on the 1D version of quantum electrodynamics, namely the Schwinger model \cite{Schwinger}. We consider its lattice regularized version using the Kogut-Susskind formulation \cite{KogutSusskind} for fermionic matter coupled to gauge fields, according to which the two components of a Dirac spinor (electron and positron) sit on even and odd sites. The corresponding Hamiltonian on an open chain of $N$ sites reads:
\begin{align}
\nonumber
H = & -  i w  \sum_{n=1}^{N-1} \big( \psi^\dagger_n e^{i \varphi_{n,n+1} } \psi_{n+1} - h. c. \big)   \\ 
	 & + J \sum_{n=0}^{N} (L_{n,n+1}+\theta/2\pi)^2  + m \sum_{n=1}^N ( - 1 )^n \psi^\dagger_n \psi^{\mathstrut}_n
\label{modelH}
\end{align}
The system can be seen as a chain of $N$ spinless fermions $\psi_n$ living on the sites and $N+1$ unbounded bosons $L_{n,n+1}$ living the on the links, the former and the latter being the matter and gauge degrees of freedom, respectively. $L$ and $\varphi$ stand for the electric field and the vector potential, respectively, and they are conjugate variables: $\left[ L , \varphi  \right]  = -i$; $\theta$ is a the lattice version of a topological angle, that we used below to tune between confined ($\theta\neq\pi$) and deconfined ($\theta=\pi$) regimes~\cite{coleman1976more}. 

The first term in the Hamiltonian represents the coupling between matter and gauge fields, the second is the electrostatic energy, and the third term gives a mass to the fermions. We will set $m=0$ in the following for the sake of simplicity (this term is not essential for the phenomenon we describe).
The Hamiltonian commutes with the generator of gauge transformations:
\begin{equation}
G_{n}  = L_{n+1,n} - L_{n,n-1} - \psi^\dagger_n \psi^{\mathstrut}_n + \frac{1}{2} [ 1 - (-1)^n ]. 
\label{gausslaw}
\end{equation}
The local symmetry generated by $G_n$ breaks the Hilbert space in superselection sectors.
States $|\Psi\rangle_{q_1, q_2, ...q_N}$ in each of those sectors are labeled by the distribution of background static charges $(q_1, q_2, ...q_N)$, defined as:
\begin{equation}
G_n |\Psi\rangle_{q_1, q_2, ...q_N} = q_n |\Psi\rangle_{q_1, q_2, ...q_N}
\label{gausslaw2}
\end{equation}
which is a discretized version of Gauss law. 
Disorder-free many-body localization dynamics in this system has been reported in Ref.~\cite{brenes2018}. There, the idea was to use superselection sectors in a clean system as an effective source of correlated disorder. Other signatures of MBL in the presence of disordered on-site potentials were reported in Ref.~\cite{Sondhi2018}. Here, instead, we study the system properties to the presence of random, static background charges, that, for the sake of simplicity, we randomly choose in the set $q_j=\{0,\pm1\}$ with equal probability. 

A computationally convenient representation is obtained via explicit integration of the gauge fields. This is a consequence of the well-known fact that Gauss law can be integrated exactly in one dimension. The mapping follows standard techniques, including a Jordan-Wigner transformation to cast the theory in spin language, and is reviewed in great detail in Ref.~\cite{Hamer1997}.
The resulting dynamics is dictated by a spin-$1/2$ chain. We define as $\sigma^\alpha_j$ the corresponding Pauli matrices at site $j$. The resulting Hamiltonian reads: 
\begin{equation}
H_0 = w H_{\mathrm{Hop}} + J H_{\mathrm{Int}} + J H_{\mathrm{Dis}}
\label{H_model}
\end{equation}
where $H_{\mathrm{Hop}} $ is just the hopping term $H_{\mathrm{Hop}}  = - \sum_{n=1}^{N-1} \big( \sigma^+_n \sigma^-_{n+1} + \text{h.c.} \big)$, while the second and third terms read
\begin{equation}
H_{\mathrm{Int}} =  \frac{1}{2} \sum_{n=1}^{N-2} \sum_{\ell = n + 1}^{N-1} \big(N-\ell \big) \sigma^z_n \sigma^z_\ell,
\end{equation}
\begin{equation}
H_{\mathrm{Dis}} = \frac{1}{2} \sum_{n=1}^{N-1} \left( \sum_{\ell = 1}^n \sigma^z_\ell \right)  \left[ 2  \sum_{j=1}^n q_j +  \frac{ (-1)^n - 1 }{2} +\frac{\theta}{\pi}\right]
\end{equation}
and describe the Coulomb interaction between dynamical charges (both terms), and the interaction between dynamical and static ones (the last term). Note that the parameter $J$ measures at the same time disorder and interaction strength. The intimate relation between these two quantities is a natural consequence of the existence of Coulomb law: in any local theory in 1D, local background charges will inevitably generate a sink (or source) of the electric field, and thus their effect on the system is tied to the gauge coupling. 

Below, we set $w=1$ and consider only static charge distributions such that $\sum_n q_n = 0$ and $q_n = 0 , \pm 1 $. We set the left boundary electric field $L_{0,1} = 0$ and restrict to charge neutrality, $\sum_n \psi^\dagger_n \psi^{\vphantom{\dagger}}_n   = 0 $. In order to avoid spurious effects close to $J=0$ due to the system becoming non-interacting, we add a next-to-nearest-neighbor interaction of the form $H_\epsilon= \epsilon \sum_{n=1}^{N-2} \sigma^z_{n} \sigma^z_{n+2}$. We set $\epsilon=0.5$, which is large enough to remove finite size-integrability effects close to $J=0$.

\begin{figure}
 \includegraphics[scale=0.38]{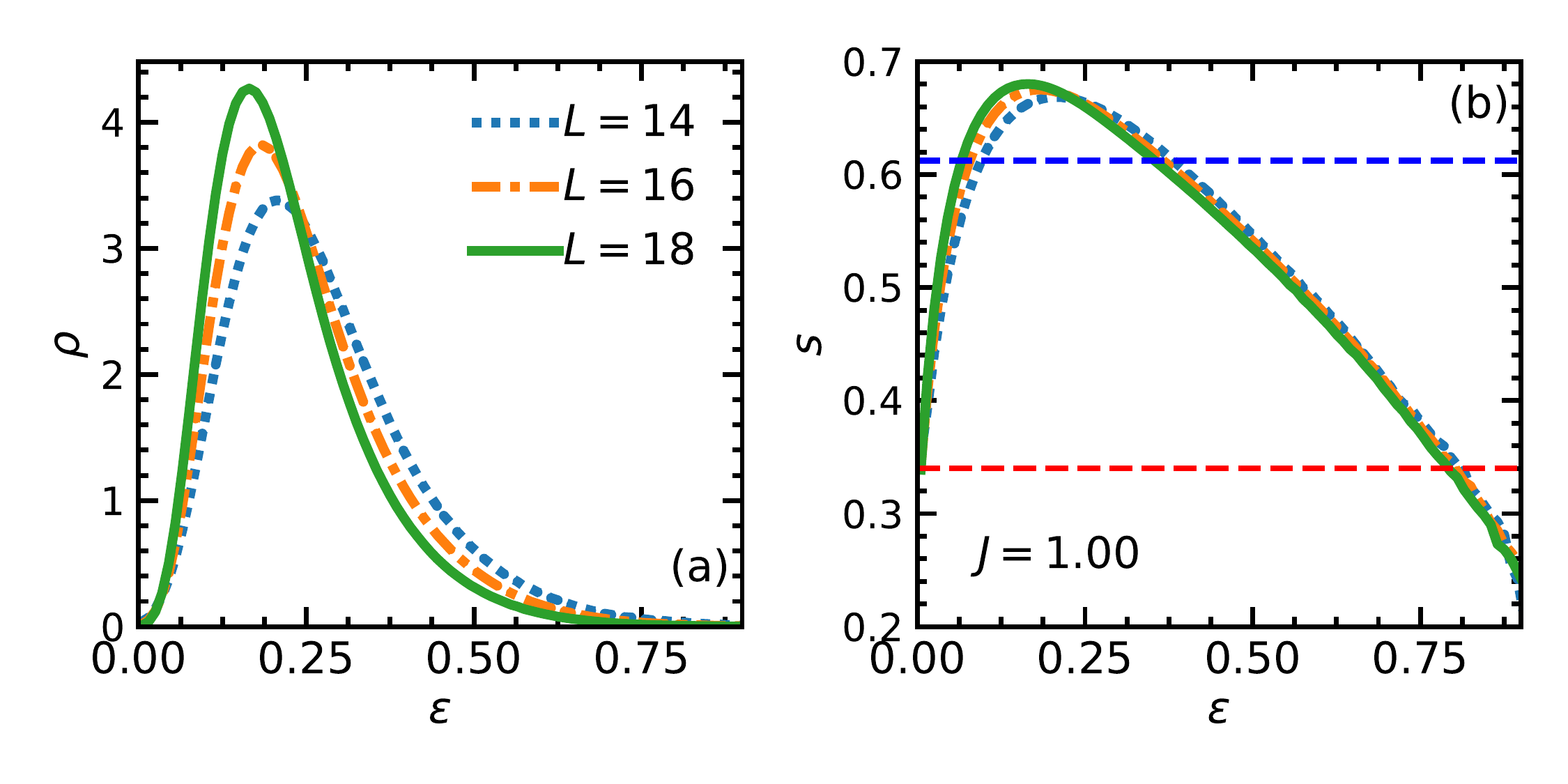}
\caption{Spectral density (a) and entropy per site (b) of the Hamiltonian Eq.~\eqref{modelH} for different $L$ and $J=1$ as a function of the rescaled energy $\varepsilon = (E-E_{\mathrm{min}})/(E_{\mathrm{max}} - E_{\mathrm{min}} )$. 
The blue dashed lines cut the spectrum keeping only the eigenvalues $E$ s.t. $s(E)/s_{\mathrm{max}} > A$. We employed $A=0.5$ for the computation of the level statistics $r$ (blue) and $A=0.9$ for the computation of the spectral form factor (red).}
 \label{fig2}
\end{figure}

\paragraph{Spectral diagnostics: average level spacing ratio. --}

To capture the ergodic to non-ergodic transition, we focus on spectral properties. We study the Hamiltonian in Eq.~\eqref{H_model} by full diagonalization in the Hilbert space sector with zero total spin along $z$. In the gauge theory picture, this means zero dynamical total charge. We define the ratio between nearby gaps as
\begin{equation}
r_\alpha = \frac{ \mathrm{Min} \{ \Delta E_\alpha  , \Delta E_{\alpha + 1 } \}  }{ \mathrm{Max} \{ \Delta E_\alpha  , \Delta E_{\alpha + 1 } \} }
\label{r_ratio}
\end{equation}
here $\alpha$ labels the eigenvalues of $H$ for a given disorder realization. We average $r$ over a spectral window centered on the most-likely eigenvalue, and over 1000 and 100 disorder realizations for $L<18$ and $L=18$ respectively. 

\begin{figure}
{\hspace*{-6mm} \includegraphics[scale=0.38]{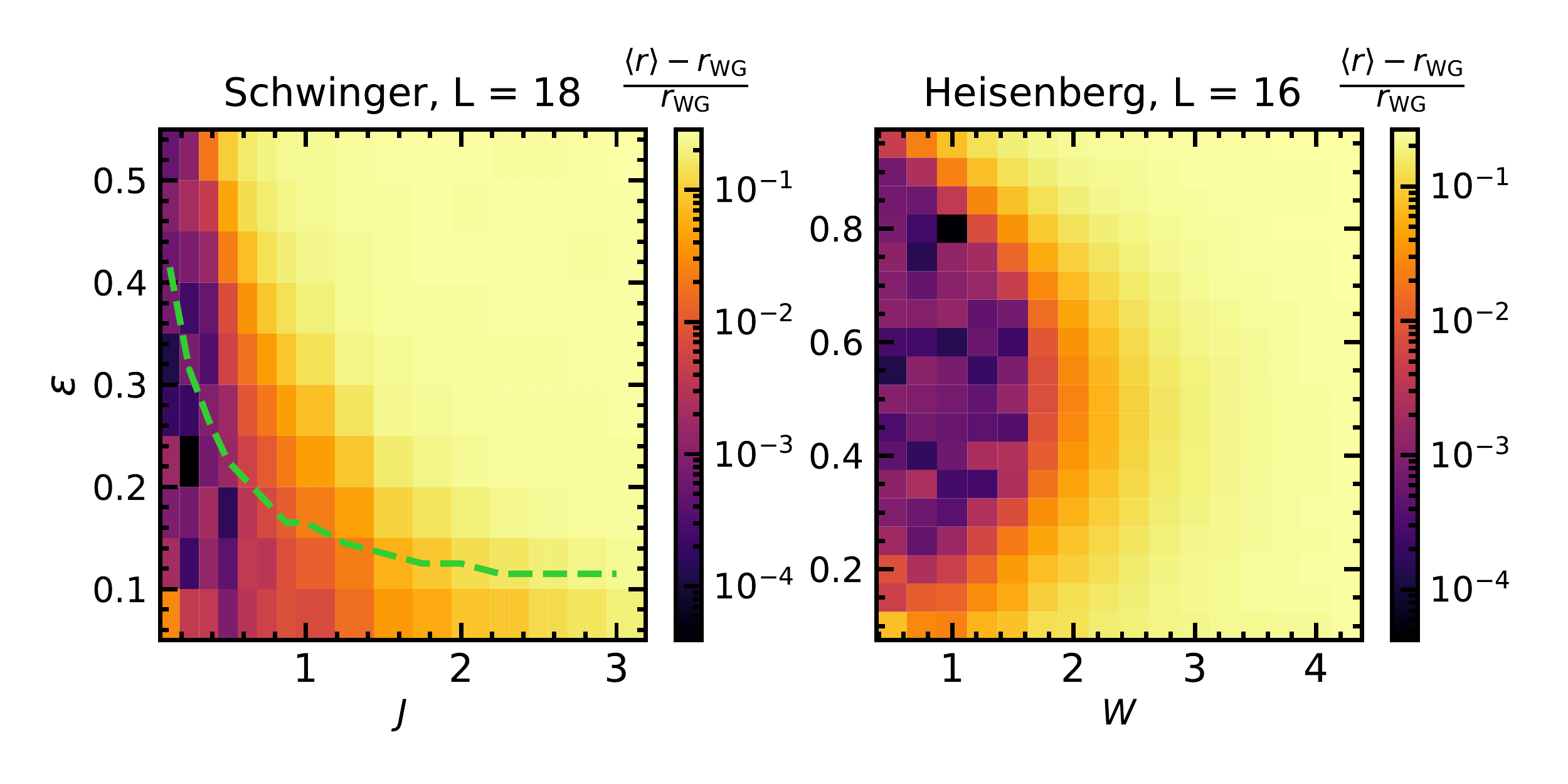}}
\caption{Energy-resolved $r$ as a function of the rescaled energy $\varepsilon$, and gauge coupling $J$ (gauge theory, left) and disorder strength $W$ (Heisenberg model, right). The green dashed line indicates the position of the maximum of the spectral density. 
}
\label{fig3} 
\end{figure}

As illustrated in Fig.~\ref{fig2} (a), the Coulomb interaction makes the eigenvalue distribution $\rho$ strongly asymmetric, due to the super-linear scaling of the largest eigenvalues in the spectrum. We thus cut the tails of the spectral density $\rho$ by monitoring the thermodynamic entropy per site: $s = \log \rho / L $. This quantity has a well defined thermodynamic limit (see Fig.~\ref{fig2} (b)) and can be used to select the most relevant part of the spectrum ensuring a smooth scaling with the system size. To compute the level statistics $r$ we keep only the eigenvalues $E$ for which $s(E)/s_{\mathrm{max}} > 0.9$ (blue dashed line in Fig.~\ref{fig2} (b)). This corresponds to a fraction of eigenstates larger than $0.4$, at $L=14$, and it increases with $L$. For gauge theories, this procedure overestimates $\langle r\rangle$ at finite size: the reason is that, differently from spin chains, states at lower energy densities are typically less affected by Coulomb law, and thus less localized, at small system sizes. This is illustrated in Fig.~\ref{fig3}(a), where the energy resolved $r$-value is plotted as a function of gauge coupling and energy density~\footnote{Similar behvior occurs in the Bose-Hubbard model~\cite{Sierant_2018}}. Considering the full spectrum does not lead to quantitative changes in the transition region.

The resulting scaling of $r$ versus $J$ is plotted in Fig.~\ref{fig1}. The results illustrate how compatibility with a Wigner-Dyson distribution of the energy levels breaks down at around $J\simeq0.5$; contrary to the Heisenberg model case (where the critical disorder strength increases by 50\% when comparing $L=12$ and $L=18$), there is no appreciable finite-volume drift. We note that this behavior is fully compatible with the energy-resolved patter of $r$ plotted in Fig.~\ref{fig3}(a): indeed, only states very close to the ground state are not localized, and as such, the global value of $r$ is dominated by the vast majority of states that is localized (note that the vertical axis in Fig.~\ref{fig3}(a) is limited to $\varepsilon\in[0.05, 0.55]$ for the sake of clarity). The ergodic region (shaded) is followed by a regime where $\langle r\rangle$ takes intermediate values: while it is not possible to reliably distinguish between emergent integrability (denoted by Poisson statistics) and an intermediate value of $r$, there is a clear finite-size trend toward the former for $J>1$. Within statistical errors, we do not observe a clear crossing: longer chains routinely have smaller $r$ values with respect to shorter chains.

\paragraph{Spectral diagnostics: form factor. --} As a further evidence of breakdown of ergodicity, we analyze spectral correlations which go beyond nearby eigenvalues via the spectral form factor (SFF), defined as 
\begin{equation}
K(\tau) = \frac{1}{\mathcal{Z}}  \left|  \sum_\alpha g( \tilde{E}_\alpha )e^{ i 2 \pi \tau \tilde{E}_\alpha }  \right|^2
\label{sff}
\end{equation}
where $ \tilde{E}_\alpha $ are the unfolded eigenvalues. In order to smooth the effects due to boundaries of the spectrum, we apply a gaussian filter $g( x ) = e^{- \frac{(x-\mu)^2}{2 (\eta \sigma)^2} }$, with $\mu$ and $\sigma$ the average and variance of the disorder realization of the unfolded spectrum. $\eta$ quantifies the strength of the filter, and we take $\eta = 0.3$ in what follows. $\mathcal{Z} =  \sum_\alpha |   g(\tilde{E}_\alpha) |^2$ is a normalization s.t. $K(\tau) \simeq 1$ for large $\tau$. Before applying the filter, we cut the edges of the spectrum according to $s(E)/s_{\mathrm{max}} > 0.5$, which means we take a fraction of eigenvalues larger than 0.9. Upon unfolding, the Heisenberg time $t_{\mathrm{H}}$, corresponding to the timescale beyond which the discrete nature of the spectrum manifest itself and thus non-universal features kick in, is set to unity. The SFF in Eq.~\eqref{sff} is computed for each disorder realization for $\tau \in [0,1]$ and an average over disorder is performed for each value of $\tau$.

\begin{figure}
{\hspace{-0.3cm}\includegraphics[scale=0.34]{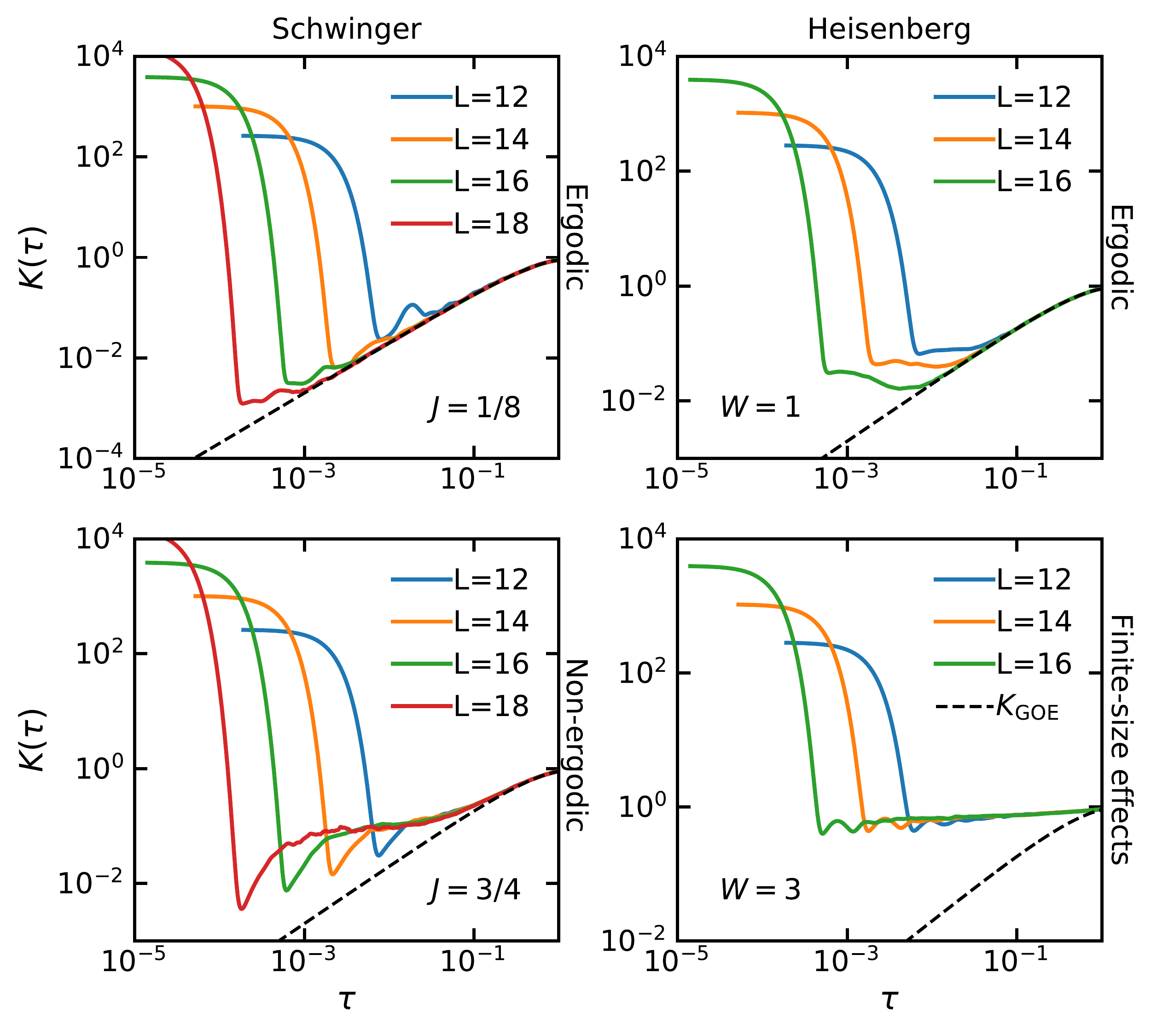}}
\caption{Comparison between the SFF of the Hamiltonian Eq.\,\eqref{H_model} (left) and of the Heisenberg model with on-site disorder (right). In the deep ergodic region (top) the SFF approaches the GOE prediction at times which decrease exponentially with the size of the system. For $J=1$ in the LGT (bottom left) the bulk of the spectrum is non-ergodic and the SFF deviates from the GOE prediction at intermediate times. For $W=3$ in the Heisenberg chain (bottom right) the level statistics is still flowing to WG, however the small effective localization length prevents accessing localization properties of thermodynamic limit. }
\label{fig4} 
\end{figure}

The analysis of $K(\tau)$ allows to probe if the system is ergodic~\cite{Serbyn_2017,untajs2019quantum,sierant2019}. This can be done by comparing the averaged SFF with the SFF expected from an ensemble of orthogonal random matrices with gaussian entries (GOE), $K_{\mathrm{GOE}} = 2 \tau - \tau \log ( 1 + 2 \tau ) $. We call $t_{\mathrm{GOE}}$ the time at which the averaged SFF approaches the GOE prediction. If the system is ergodic~\cite{sierant2019}, this corresponds to the Thouless time, and one has $t_{\mathrm{GOE}}/t_{\mathrm{H}} \to 0$ in the thermodynamic limit (specifically, the Thouless time shall increase algebraically with $L$).

In Fig.~\ref{fig4}(a,b), we plot the spectral form factor in the Schwinger model and Heisenberg model in their ergodic regions: in both cases, the Thouless time is clearly decreasing with system size, further confirming the ergodic nature of the phase. The results in Fig.~\ref{fig4}(c) correspond to a regime of gauge couplings whose $r$ value departs from GOE: such departure is indeed confirmed by the fact that the $t_{\mathrm{GOE}}/t_{\mathrm{H}}$ is not decreasing with system size, and oppositely, the SFF seems to collapse on a finite linear region, which implies $\ln t_{\mathrm{GOE}}\sim L$; this timescale directly indicates that the system is not ergodic, and it is suggestive of an emergent localization even at this value of the coupling (even if a direct connection between $t_{\mathrm{GOE}}$ and transport properties is established, to the best of our knowledge, only under the assumption of ergodicity). We note that, in this parameter regime, we do not observe saturation of the Thouless time, which is instead evident in spin models (see Fig.~\ref{fig4}(d) and Ref.~\cite{untajs2019quantum}).

\begin{figure}
\begin{center} \includegraphics[scale=0.34]{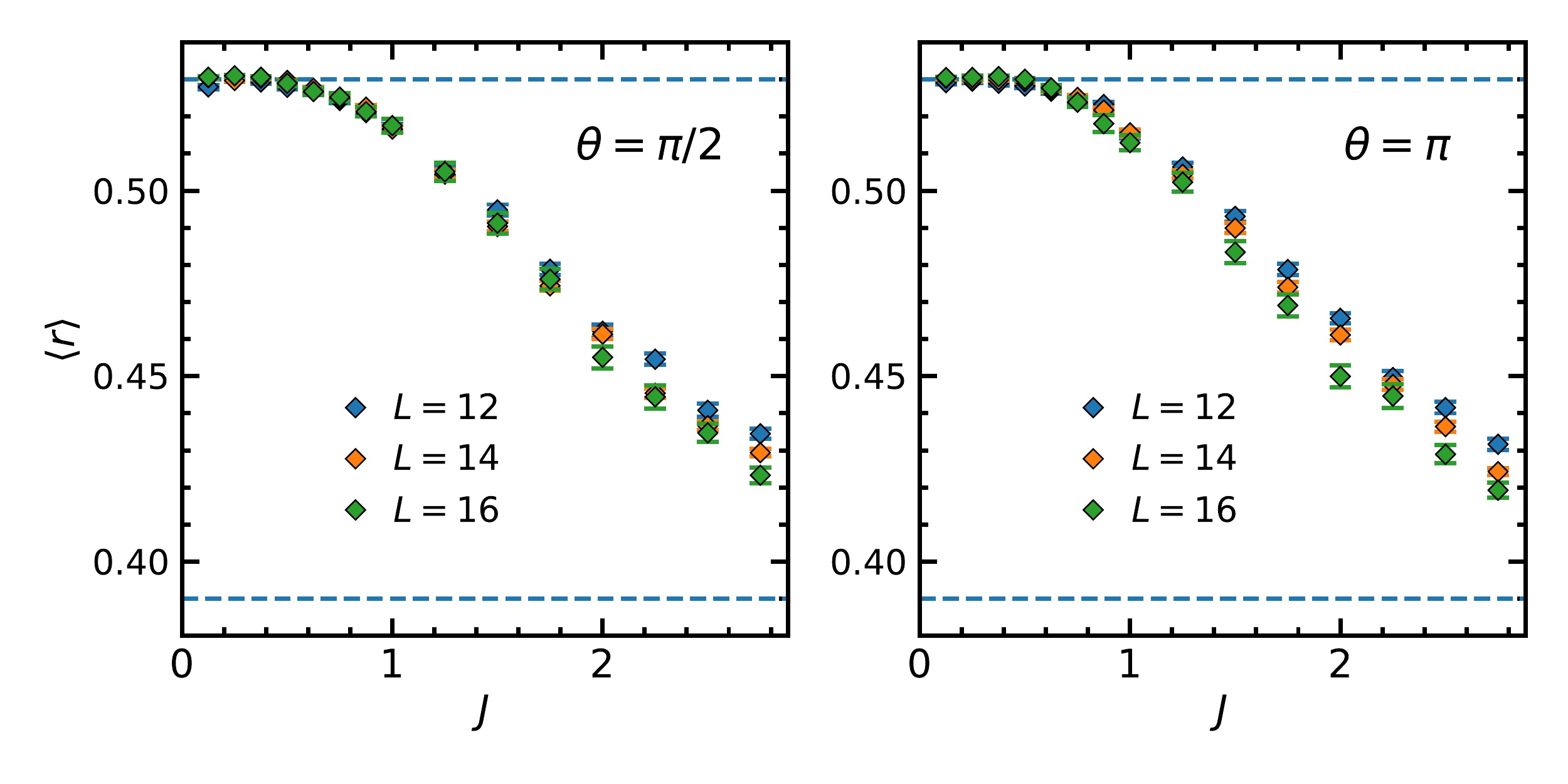}
\end{center} \vspace{-6mm}
\includegraphics[scale=0.34]{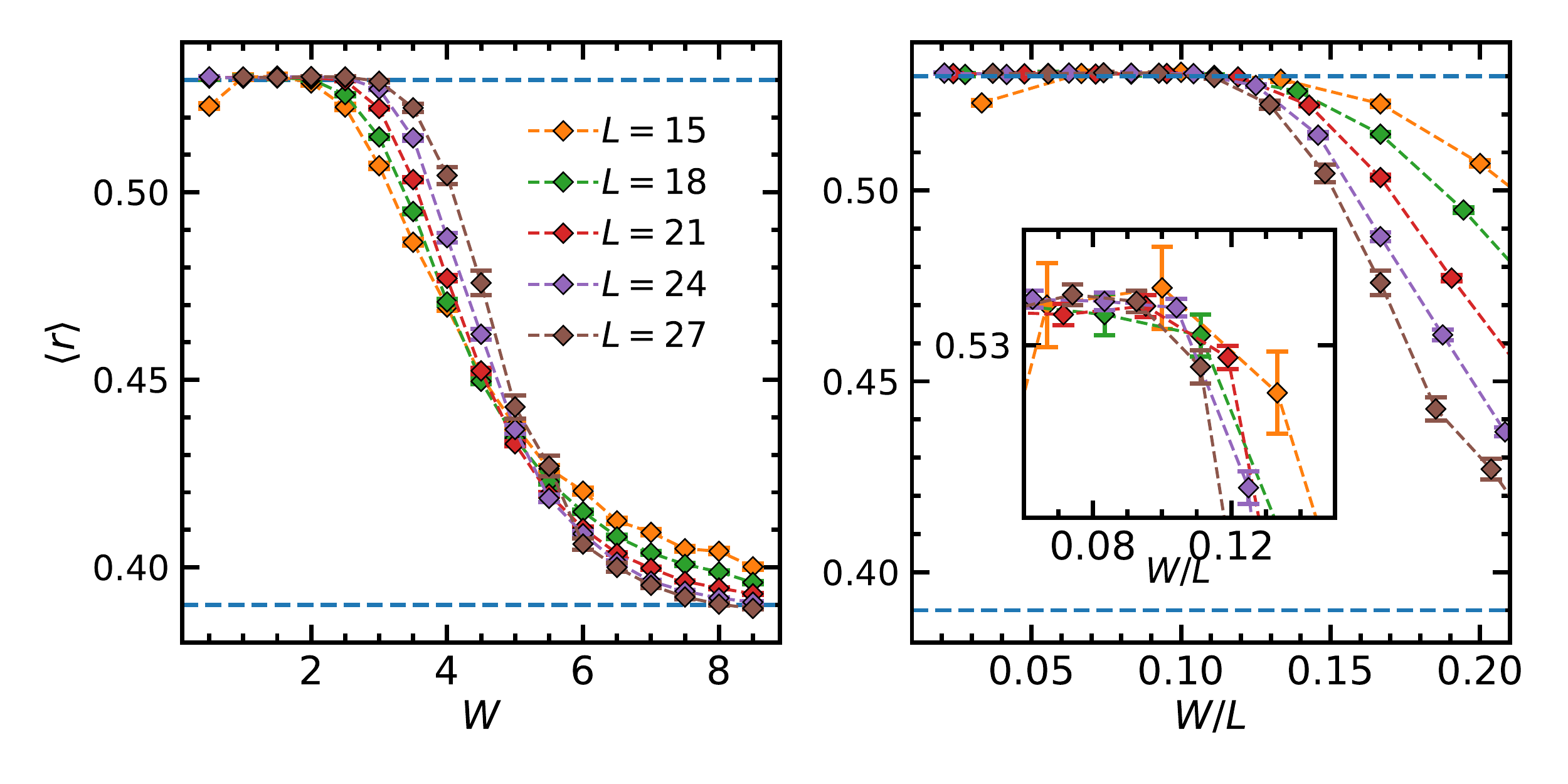} 
\caption{(Top row) Average level spacing ratio for the Schwinger model Eq.\,\ref{modelH} with non-zero topological angle $\theta$. A non-zero $\theta$ does not change the outcome w.r.t. $\theta=0$ (see Fig.\,\ref{fig1} (b)), not even in the deconfined regime $\theta = \pi$.  (Bottom row) Average level spacing ratio for the constrained spin model $H_{QLM}$, corresponding to the Schwinger model with truncated gauge fields. The finite-size scaling of $\langle r \rangle$ exhibits the same phenomenology as in the Heisenberg chain: $\langle r \rangle$ vs $W$ (left) shows a crossing point drifting on the right for increasing $L$, $\langle r \rangle$ vs $W/L$ (right) gives a good data collapse for $W/L < 0.1$.}
\label{fig5} 
\end{figure}

\paragraph{Origin of ergodicity breaking. -} As discussed in the introduction, we conjecture that the origin of ergodicity breaking in lattice gauge theories stems from the fact that Coulomb law - which is acting at all energy scales - further constrains the system dynamics, and thus acts as an amplifier of any background disorder. 
In fact, for increasing system sizes, a larger fraction of the states of the spectrum will feature regions with a large accumulation of charge: as a consequence, the electrostatic energy (which is locally unbounded) becomes dominant and the effect of Coulomb interactions is enhanced.
The presence of an unbounded energy density, which contrasts with the usual behaviour of spin models, does not affect low-energy states, but has important consequences on the rest of the spectrum: for instance, it systematically reduces the number of available resonances when size is increased. 
In order to substantiate this statement, we studied (1) the Schwinger model in its deconfined regime, $\theta=\pi$, and (2) a quantum link version of the model with truncated gauge fields, where Coulomb law is washed out by the truncation. We note that the fact that (de)confinement is not crucial here is not unexpected, as the latter is a phenomenon that only dictates the dynamics in the vicinity of the vacuum state. 

In Fig.~\ref{fig5}(a-b), we show $r$ versus $J$ for two other values of the topological angle in the Schwinger model. Within error bars, we do not observe any difference between confining and deconfining regimes: in both cases, ergodicity breaks down in the same coupling window. 

In Fig.~\ref{fig5}(c-d), we instead show $r$ versus the disorder strength $W$ in a quantum link model in the presence of a background disorder (see Ref.~\cite{supmat} for details on the model)~\cite{banerjee2012atomic}. In the presence of strong nearest-neighbor interactions, the system dynamics can be mapped exactly~\cite{surace2019lattice,supmat} into a constrained spin model of the form $H_{QLM} = \sum_i (W_i n_i - \sigma_i^x )$, $n_i n_{i+1} = n_{i} n_{i+2} = 0$, where $n=( 1-\sigma^z )/2 $. We have tried a collapse scaling following ~\cite{alet2015}, and assuming finite transition point $W_c$ and correlation length critical exponent $\nu$. The best fitting $W_c$ and $\nu$ seem to increase linearly with size. The scaling or $r$ follows rather closely the functional form proposed in Ref.~\cite{untajs2019quantum}. These two observations indicate that, even in this model, the available system sizes are not sufficient to determine whether ergodicity is broken in the thermodynamic limit~\cite{p2019study}. Overall, the findings on these two models support our conjecture above.

\paragraph{Conclusions and outlook. --} We have provided numerical evidence for the breakdown of ergodicity in disorderd $U(1)$ lattice gauge theories. Our results do not immediately indicate if localization kicks in right after such a breakdown, or if an intermediate non-ergodic, delocalized regime occurs. Further studies based on localication-specific diagnostics may elucidate this aspect. The dynamical consequences of our results are immediately testable on quantum simulation platforms, where many-body dynamics of $U(1)$ lattice gauge theories has been recently realized~\cite{martinez2016real,Bernien2017,surace2019lattice}, and, based on the nature of the interactions, might be extended to Yang-Mills theories.

\paragraph{Acknowledgements.}

We thank D. Abanin, M. Amini, J. Chalker, M. Heyl, V. Kravtsov, D. Luitz,  F. Pollmann, T. Prosen and J. Zakrzewski for discussions. This work is partly supported by the ERC under grant number 758329 (AGEnTh), by the Quantera programme QTFLAG, and has received funding from the European Union's Horizon 2020 research and innovation programme under grant agreement No 817482. This work has been carried out within the activities of TQT. 

\bibliography{MBLbib.bib}
\end{document}


\title{Supplementary material -- Breakdown of ergodicity in disordered $U(1)$ lattice gauge theories}
\author{G. Giudici}
\affiliation{The Abdus Salam International Center for Theoretical Physics, Strada Costiera  11,  34151  Trieste,  Italy}
\affiliation{SISSA, via Bonomea 265, 34136 Trieste, Italy}
\author{F. M. Surace}
\affiliation{The Abdus Salam International Center for Theoretical Physics, Strada Costiera  11,  34151  Trieste,  Italy}
\affiliation{SISSA, via Bonomea 265, 34136 Trieste, Italy}
\author{J. E. Ebot}
\affiliation{The Abdus Salam International Center for Theoretical Physics, Strada Costiera  11,  34151  Trieste,  Italy}
\author{A. Scardicchio}
\affiliation{The Abdus Salam International Center for Theoretical Physics, Strada Costiera  11,  34151  Trieste,  Italy}
\affiliation{INFN Sezione di Trieste, Via Valerio 2, 34127 Trieste, Italy}
\author{M. Dalmonte}
\affiliation{The Abdus Salam International Center for Theoretical Physics, Strada Costiera  11,  34151  Trieste,  Italy}
\affiliation{SISSA, via Bonomea 265, 34136 Trieste, Italy}

\date{\today}


\maketitle

\paragraph{Quantum Link Model -- }
In this section we introduce an alternative model of Quantum Electrodynamics with respect to the lattice Schwinger model in Eq.(1) of the main text. The Hamiltonian has a similar structure
\begin{multline}
   H_{QLM}=-w\sum_{j=1}^{L-1}({\psi}^\dag_j  U^{\mathstrut}_{j,j+1}{\psi}^{\mathstrut}_{j+1} + \textnormal{h.c.})+ J\sum_{j=1}^{L-1} E_{j,j+1}^{\mathstrut 2 }\\
   + m \sum_{j=1}^L(-1)^j {\psi}_j^\dag{\psi}^{\mathstrut}_j, \label{QLM}
\end{multline}
where the first term is the matter-gauge coupling, the second is the
electrostatic energy, and the third is the mass term. The gauge field operators are here represented by spin matrices
$ E_{j, j+1}= S^z_{j, j+1}$ and  $ U_{j, j+1}= S^+_{j,j+1}$, such that the commutation relation $[ E_{j,j+1}, U_{j,j+1}]= U_{j,j+1}$ is satisfied. The Hamiltonian commutes with the local generators defined as
 \begin{equation}
G_j= E_{j,j+1}- E_{j-1,j}-{\psi}_j^\dagger {\psi}^{\mathstrut}_j+\frac{1-(-1)^j}{2}.
\label{eq:Gauss}
\end{equation}
This model resembles the lattice Schwinger model in Eq.(1) of the main text, but its Hilbert space is smaller: on each site the gauge field is truncated to a maximal value given by the dimension of the spin representation. Since the electrostatic term suppresses the large values of $E_{j,j+1}^2$, the truncation has minor effects on the ground state properties. The properties of the full spectrum, on the other hand, are expected to change drastically. 
The model in Eq.~\ref{QLM} in the gauge invariant sector (i.e. the one with no static charges, where $G_j=0$ for every $j$) can be mapped to a constrained spin $1/2$ model (the PXP model), as discussed in reference \onlinecite{surace2019lattice}.

The Hamiltonian Eq.\,\eqref{QLM} is mapped to $H_{QLM} = \sum_i (-2m~ n_i - w\sigma_i^x )$,
 with $n=( 1-\sigma^z )/2 $, and it is restricted to the Hilbert space with $n_i n_{i+1} = 0$. In order to have a model where larger system sizes are accessible, we further constrain the Hilbert space to the sector with $n_in_{i+2}=0$: this constraint emerges when we include a strong next-to-nearest-neighbour interactions between electric fields in Eq.~\eqref{QLM}. The disorder is introduced by making the coefficient $W=-2m$ site-dependent. We obtain the Hamiltonian 
 \begin{equation}
     H_{QLM} = \sum_i (W_i n_i - \sigma_i^x )\hspace{0.6cm} n_i n_{i+1} = 0,\, n_i n_{i+2} = 0
 \end{equation}
where we have fixed $w=1$, and the $W_i$ are drawn from a uniform distribution in the interval $[-W, W]$.
 
\bibliography{MBLbib.bib}